\documentclass[onecolumn]{aastex6}

\collaborationName{Friends of AASTeX}

\begin{document}


\title{Quasi-periodic fast propagating magnetoacoustic waves during the magnetic reconnection between solar coronal loops}


\author{Leping Li\altaffilmark{1,2,3}, Jun Zhang\altaffilmark{1,3}, Hardi Peter\altaffilmark{4}, Lakshmi Pradeep Chitta\altaffilmark{4}, Jiangtao Su\altaffilmark{1,3}, Hongqiang Song\altaffilmark{2}, Chun Xia\altaffilmark{5,6}, and Yijun Hou\altaffilmark{1,3}}


\altaffiltext{1}{CAS Key Laboratory of Solar Activity, National Astronomical Observatories, Chinese Academy of Sciences, Beijing 100101, People's Republic of China}
\altaffiltext{2}{Shandong Provincial Key Laboratory of Optical Astronomy and Solar-Terrestrial Environment, and Institute of Space Sciences, Shandong University, Weihai, Shandong 264209, People's Republic of China}
\altaffiltext{3}{University of Chinese Academy of Sciences, Beijing 100049, People's Republic of China}
\altaffiltext{4}{Max-Planck Institute for Solar System Research, D-37077 G\"{o}ttingen, Germany}
\altaffiltext{5}{School of Physics and Astronomy, Yunnan University, Kunming 650050, People's Republic of China}
\altaffiltext{6}{Centre for mathematical Plasma Astrophysics, Department of Mathematics, KU Leuven, Celestijnenelaan 200B, B-3001 Leuven, Belgium}



\begin{abstract}

Employing Solar Dynamics Observatory/Atmospheric Imaging Assembly (AIA) multi-wavelength images, we have presented coronal condensations caused by magnetic reconnection between a system of open and closed solar coronal loops. In this Letter, we report the quasi-periodic fast magnetoacoustic waves propagating away from the reconnection region upward across the higher-lying open loops during the reconnection process. On 2012 January 19, reconnection between the higher-lying open loops and lower-lying closed loops took place, and two sets of newly reconnected loops formed. Thereafter, cooling and condensations of coronal plasma occurred in the magnetic dip region of higher-lying open loops. During the reconnection process, disturbances originating from the reconnection region propagate upward across the magnetic dip region of higher-lying loops with the mean speed and mean speed amplitude of 200 and 30 km\,s$^{-1}$, respectively. The mean speed of the propagating disturbances decreases from $\sim$230 km\,s$^{-1}$ to $\sim$150 km\,s$^{-1}$ during the coronal condensation process, and then increases to $\sim$220 km\,s$^{-1}$. This temporal evolution of the mean speed anti-correlates with the light curves of the AIA 131 and 304\,\AA~channels that show the cooling and condensation process of coronal plasma. Furthermore, the propagating disturbances appear quasi-periodically with a peak period of 4 minutes. Our results suggest that the disturbances represent the quasi-periodic fast propagating magnetoacoustic (QFPM) waves originating from the magnetic reconnection between coronal loops.

\end{abstract}




\section{Introduction} \label{sec:introduction}

Magnetic reconnection plays an elemental role in magnetized plasma systems, e.g., solar corona and planetary magnetospheres, throughout the universe. It shows the re-configuration of magnetic field geometry, and is used to explain the release of magnetic energy and its conversion to other forms, e.g., thermal, kinetic, and acceleration of non-thermal particles \citep{asch02}. The reconnection process is difficult to observe directly. However, because the magnetic flux is frozen into plasma in the solar corona, the coronal structures, e.g., coronal loops, and their evolution usually outline the magnetic field geometry and its change. Employing remote sensing data, many observational evidences of reconnection have been presented in solar physics, such as X-type structures \citep{su13,li16b}, reconnection inflows \citep{yoko01,huang18}, and outflows \citep{taka12,li16a,ning16}, current sheets \citep{lin05,li16c,xue18}, plasmoid ejections \citep{liu10,tian14,cheng18}, loop-top hard X-ray sources \citep{masu94,li09}, cusp-shaped post-flare loops \citep{tsun92,yan18}, and supra-arcade downflows \citep{mcke00,sava11}.

Numerous theoretical studies of reconnection have been undertaken to explain solar flares \citep{shib99,prie00}, that rapidly release the magnetic energy. During the flares, observations of quasi-periodic fast propagating magnetoacoustic (QFPM) waves have been reported \citep{liu12,shen13,nist14}. Using Atmospheric Imaging Assembly \citep[AIA;][]{lemen12} images onboard the Solar Dynamics Observatory \citep[SDO;][]{pesn12}, \citet{liu11} presented the QFPM wave trains emanating near a flare kernel, and found the strongest signal with the 181\,s period temporally coincides with the flare quasi-periodic pulsations. In another flare, \citet{shen12} noted almost all the flare main frequencies are consistent with those of the QFPM waves. QFPM wave trains associated with flaring radio emissions have also been reported \citep{yuan13,kumar17}. All these results suggest the QFPM waves and flares are possibly excited by a common physical origin, i.e., the reconnection periodically releasing magnetic energy \citep{liu11,shen12,kumar17}. However, no reconnection process is observed during these flares. On the other hand, QFPM waves driven by reconnection have been theoretically predicted and simulated \citep{ofman11,jeli12,taka16}.  Nevertheless, QFPM waves associated with reconnection directly are rarely observed.

Using AIA multi-wavelength images on 2012 January 19, reconnection between two sets of loops, and cooling and condensation of coronal plasma in the magnetic dip (MD) of higher-lying open loops have been reported \citep{li18a}. In this Letter, we reveal the presence of QFPM waves, originating from the reconnection region and propagating across the higher-lying loops upward. 
The observations and results are separately described in Sections\,\ref{sec:observation} and \ref{sec:results}, and a summary and discussion is given in Section\,\ref{sec:discussion}.

\section{Observations} \label{sec:observation}

SDO/AIA is a set of normal-incidence imaging telescopes, obtaining images of the solar atmosphere at 10 wavelength channels. Different AIA channels show plasma with different characteristic temperatures, e.g., 304 \AA~peaks at $\sim$0.05 MK (He II), 131 \AA~peaks at $\sim$0.6 MK (Fe VIII) and $\sim$10 MK (Fe XXI), and 171\,\AA~peaks at $\sim$0.9 MK (Fe IX). In this study, AIA 171, 131, and 304\,\AA~images, with spatial sampling and time cadence of 0.\arcsec6\,pixel$^{-1}$ and 12\,s, are employed to study the QFPM waves and  the coronal condensations during the reconnection process.

\section{Results} \label{sec:results}

\begin{figure}[ht!]
\plotone{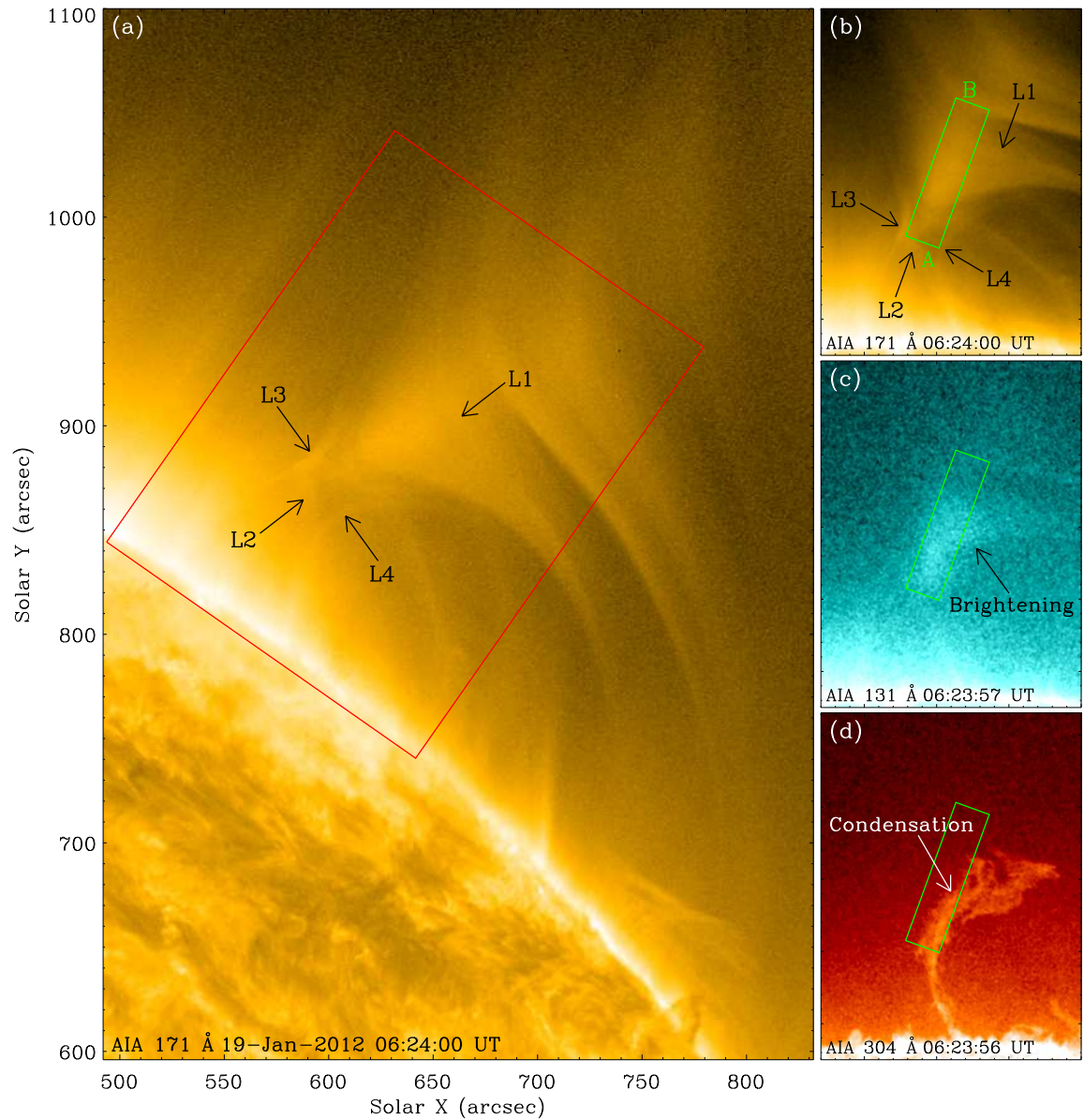}
\caption{Reconnection between coronal loops and condensations of coronal plasma. AIA 171 (a)-(b), 131 (c), and 304 \AA~(d) images. The red rectangle in (a) represents the field of view (FOV) of (b)-(d). The green rectangle AB in (b) indicates the position of the time slice of AIA 171\,\AA~images showed in Figure\,\ref{fig:measurements}(a). The green rectangles in (b)-(d) mark the regions for the light curves of the AIA 171, 131, and 304 \AA~channels displayed in Figure\,\ref{fig:condensation} separately  by red, green, and blue lines. The animation covers 8\,hr starting at 01:00 UT in 2012 January 19. (An animation of Figures\,\ref{fig:reconnection}(b)-(d) is available.)
\label{fig:reconnection}}
\end{figure}

On 2012 January 19, a set of open curved loops, L1, located above the northwestern solar limb, were observed in AIA 171\,\AA~images, see Figure\,\ref{fig:reconnection}(a). From $\sim$01:00 UT, the loops L1 moved toward the southeast, and reconnected with the lower-lying closed loops, L2. Two sets of newly reconnected loops, L3 and L4, formed and retracted away from the reconnection region. Due to the downward motion,  a MD of loops L1 forms. Cooling and condensations of coronal plasma in the MD then take place. Since the reconnection site moved to the southwest with a displacement of $\sim$35 Mm after $\sim$09:00 UT, the reconnection process from 01:00 to 09:00 UT in the region enclosed by a red rectangle in Figure\,\ref{fig:reconnection}(a) is chosen and investigated in detail. To better display evolution of the reconnection, the AIA images are rotated counter-clockwise by an angle of 35$^{\circ}$. The portion of the limb in the region of interest, i.e., the red rectangle in Figure\,\ref{fig:reconnection}(a), is thus roughly horizontal in the AIA images presented here, see Figure\,\ref{fig:reconnection}(b). AIA 131 and 304\,\AA~images are illustrated in Figures\,\ref{fig:reconnection}(c)-(d) to show the cooling and condensations of coronal plasma in the MD of loops L1 during the reconnection (see the online animated version of Figures\,\ref{fig:reconnection}(b)-(d)).  

\begin{figure}[ht!]
\plotone{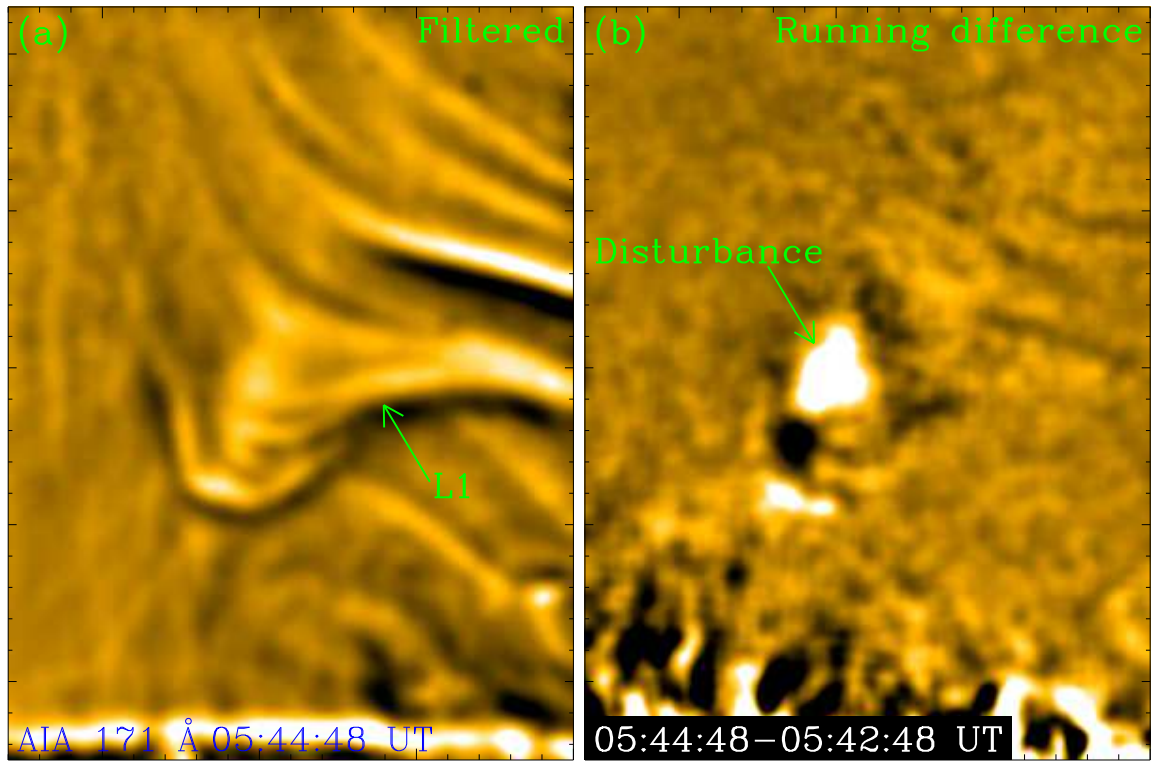}
\caption{Disturbances propagating from the reconnection region upward across the higher-lying loops. AIA 171\,\AA~filtered (a) and running difference (b) images, with the same FOV as in Figures\,\ref{fig:reconnection}(b)-(d). The animation covers 5\,hr starting at 03:30 UT. (An animation of this figure is available.)
\label{fig:disturbances}}
\end{figure}

\begin{figure}
\plotone{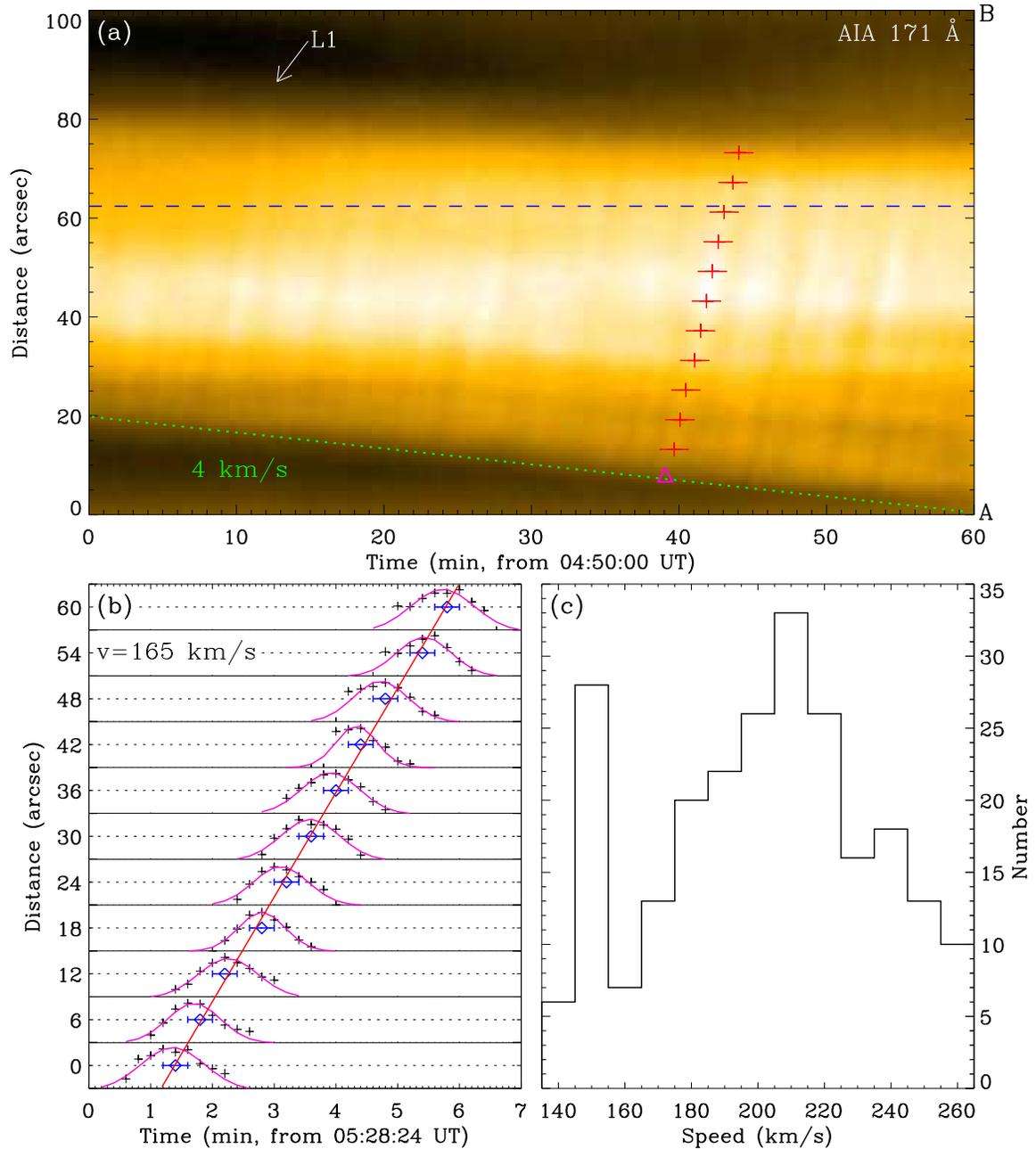}
\caption{Measurements of the parameters of disturbances. (a) A time slice of AIA 171\,\AA~images along the AB direction in the green rectangle in Figure\,\ref{fig:reconnection}(b). (b) Intensities (black pluses) of the AIA 171\,\AA~channel at positions marked by red pluses along the disturbance in (a). (c) Histogram of the speeds of disturbances. In (a), the green dotted line outlines the downward motion of loops L1 at a mean speed of 4 km\,s$^{-1}$. The blue dashed line shows the position for the intensity profile of the AIA 171\,\AA~channel displayed in Figure\,\ref{fig:wavelet}(a) by the black curve. The pink triangle marks the beginning time of the disturbance. In (b), the pink curves represent the single Gaussian profiles to the intensities. The blue diamonds and lines mark the peak positions and their error bars of the single Gaussian profiles. The red line shows the linear fit to the peak positions. 
\label{fig:measurements}}
\end{figure}

During the reconnection process, disturbances are generated from the reconnection region. Figures\,\ref{fig:disturbances}(a)-(b) show the filtered and running difference AIA 171\,\AA~images illustrating the disturbances (see the online animated version of Figure\,\ref{fig:disturbances}). After the appearance, the disturbances propagate away from the reconnection region and upward across the higher-lying loops. Along the propagating direction AB in the green rectangle in Figure\,\ref{fig:reconnection}(b), a time-slice of AIA 171\,\AA~images is made and displayed in Figure\,\ref{fig:measurements}(a). The green dotted line outlines the downward motion of loops L1, with speeds of 2-9 km\,s$^{-1}$. The steep, recurrent stripes show the propagating disturbances across the loops L1. The stripe, marked by red pluses, is chosen to display measurements of the beginning time and propagating speed of disturbances. The chosen disturbance appears from $\sim$05:29:12 UT, denoted by the pink triangle in Figure\,\ref{fig:measurements}(a). The intensities of the AIA 171\,\AA~channel at different positions along the disturbance, marked by the red pluses in Figure\,\ref{fig:measurements}(a), are plotted in Figure\,\ref{fig:measurements}(b) as black pluses. The pink curves in Figure\,\ref{fig:measurements}(b) show the single Gaussian profiles to the pluses. The blue diamonds mark peak positions of the single Gaussian profiles, with 12\,s error bars. The red line shows the linear fit to the diamonds, indicating the disturbance propagates at a constant speed of 165 km\,s$^{-1}$. For all the disturbances, each one propagates with a nearly constant speed.

\begin{figure}
\plotone{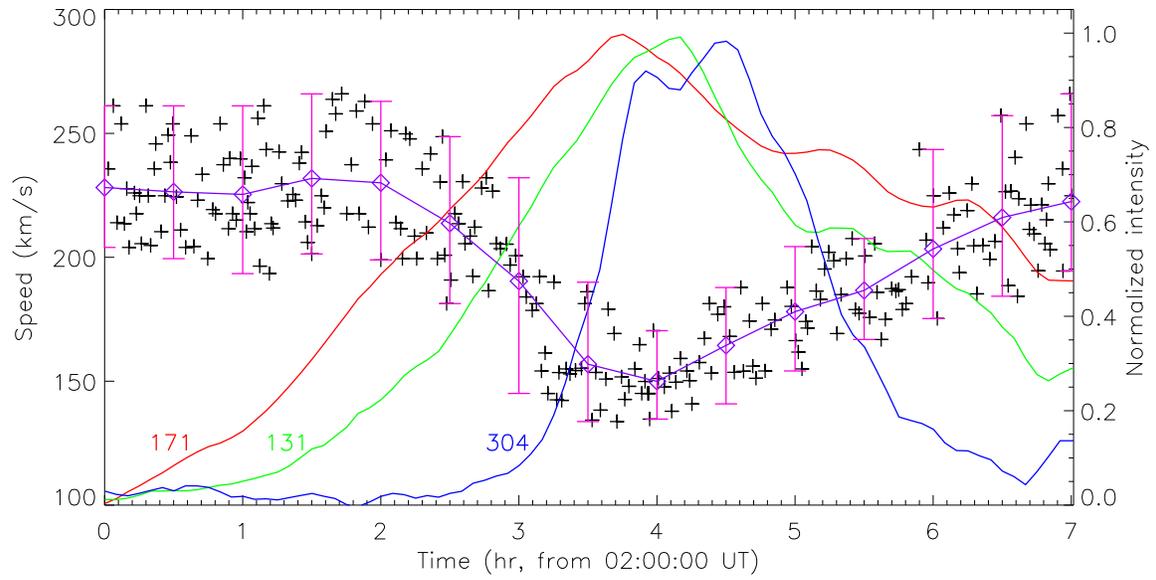}
\caption{Temporal evolution of the speeds of disturbances. The black pluses show the speeds. The purple diamonds and curve indicate the mean speeds averaged in 30 minutes. The pink vertical lines denote the speed amplitudes. The red, green, and blue curves show the light curves of the AIA 171, 131, and 304\,\AA~channels in the green rectangles in Figures\,\ref{fig:reconnection}(b)-(d). \label{fig:condensation}}
\end{figure}

Histogram of the speeds of disturbances is displayed in Figure\,\ref{fig:measurements}(c). The speeds range from 135 to 265 km\,s$^{-1}$ with a mean value of 200 km\,s$^{-1}$. Two peaks are identified at 150 and 210 km\,s$^{-1}$, respectively. The temporal evolution of the speeds is illustrated in Figure\,\ref{fig:condensation} by black pluses. The purple diamonds, ranging in 150-230 km\,s$^{-1}$ with a mean value of 200 km\,s$^{-1}$, represent the mean speeds averaged in 30\,minutes. The pink vertical lines denote the speed amplitudes, with a mean value of 30 km\,s$^{-1}$ in the range of 15-45 km\,s$^{-1}$. The mean speed remains almost constant at $\sim$230 km\,s$^{-1}$, then decreases from $\sim$04:00 UT, and reaches the valley of $\sim$150 km\,s$^{-1}$ at $\sim$06:00 UT. Thereafter, it increases slowly, and goes back to $\sim$220 km\,s$^{-1}$ at $\sim$09:00 UT. 

In the propagating path of disturbances, denoted by green rectangles in Figures\,\ref{fig:reconnection}(b)-(d), light curves of the AIA 171, 131, and 304\,\AA~channels are measured and showed in Figure\,\ref{fig:condensation} as red, green, and blue curves, respectively. All the light curves increase first, reach the peaks, and then decrease slowly. However, they peak at different times. The AIA 171\,\AA~light curve peaks at 05:45 UT, but the 131 (304)\,\AA~light curve peaks at 06:05 (06:30) UT, 20 (45) minutes later. Because no associated bright emission is observed in AIA channels showing higher-temperature plasma, e.g., 193, 211, 335, and 94\,\AA, the enhancement in the AIA 131\,\AA~light curve should originate from cooler ($\sim$0.6 MK) plasma rather than from hotter ($\sim$10 MK) regions. The AIA 171, 131, and 304\,\AA~light curves thus clearly represent the cooling and condensation process of hotter plasma in the MD of loops L1 (see the online animated version of Figures\,\ref{fig:reconnection}(b)-(d)). The plasma cools down from $\sim$0.9 MK, the characteristic temperature of AIA 171\,\AA~channel, to $\sim$0.6 MK, the characteristic temperature of AIA 131\,\AA~channel, in $\sim$20 minutes, and then to $\sim$0.05 MK, the characteristic temperature of AIA 304\,\AA~channel, in another $\sim$25 minutes. Figure\,\ref{fig:condensation} indicates the temporal evolution of the disturbance speeds anti-correlates with the AIA 131 and 304\,\AA~light curves. It means the speed decreases during the condensation process. The smaller and larger peaks of the histogram of speeds in Figure\,\ref{fig:measurements}(c) thus correspond to the disturbances within and without the condensation process, respectively.

\begin{figure}
\plotone{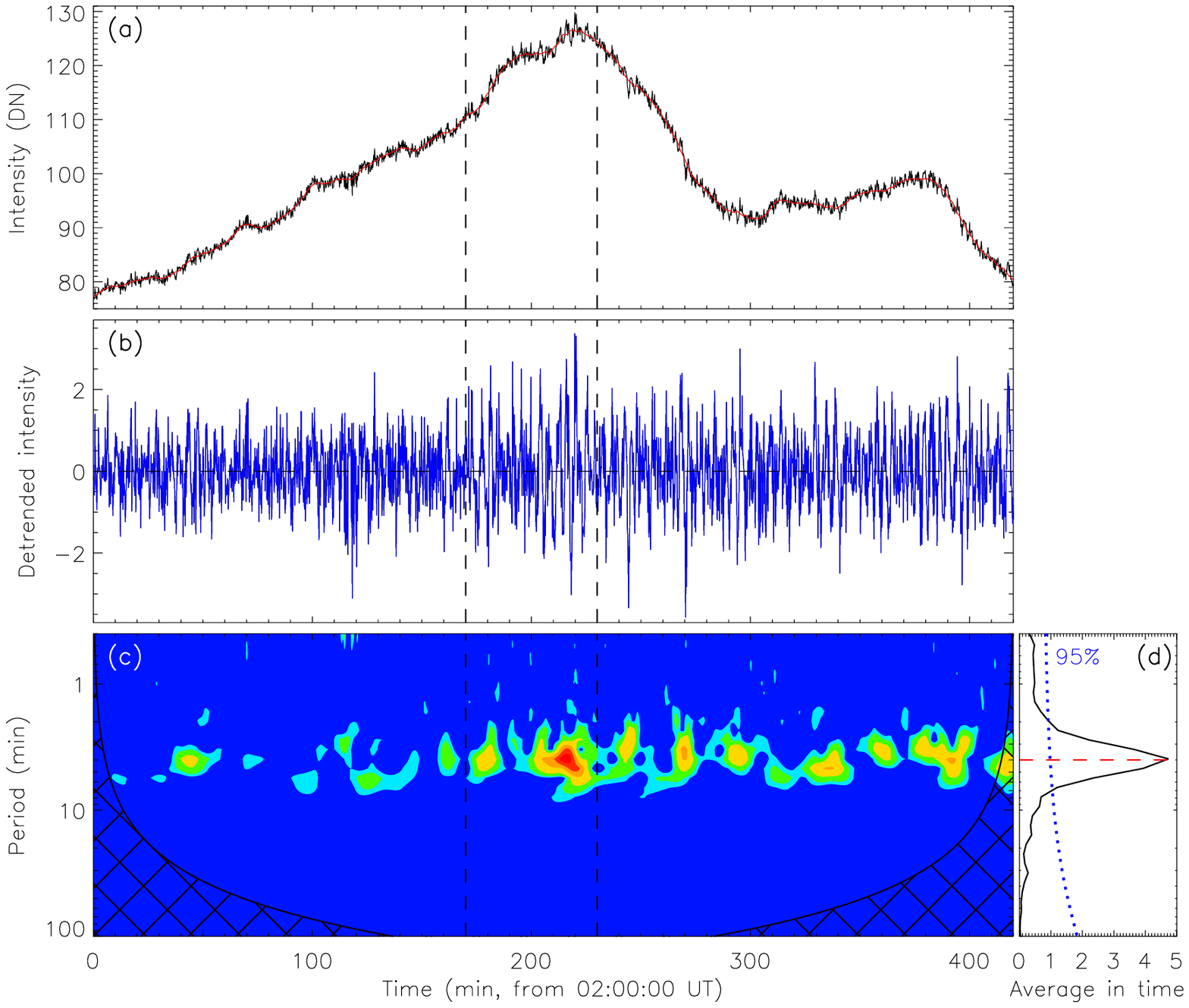}
\caption{Wavelet analysis of the intensity profile of the AIA 171\,\AA~channel. (a) Intensity profile (black curve) of the AIA 171\,\AA~channel along the blue dashed line, spatially averaged over 3 pixels, in Figure\,\ref{fig:measurements}(a), and its smoothed intensity profile (red curve) using a 6-minute boxcar. (b) Detrended intensity profile (blue curve) obtained from the intensity profile in (a) by subtracting its smoothed intensity profile. (c)-(d) Wavelet power spectrum and global wavelet power spectrum for the detrended intensity profile in (b) employing Gaussian (morlet) wavelet. The dashed vertical lines in (a)-(c) enclose the time duration of the time slice in Figure\,\ref{fig:measurements}(a). The cross-hatched regions in (c) indicate the ``core of influence" where edge affects the power. The blue dotted and red dashed lines in (d) separately mark the 95\% confidence level and the 4-minute peak period.
\label{fig:wavelet}}
\end{figure}

The disturbances recurrently originate from the reconnection region. Along the blue dashed line in Figure\,\ref{fig:measurements}(a), the intensity profile of the AIA 171\,\AA~channel, spatially averaged over 3 pixels, is calculated and shown in Figure\,\ref{fig:wavelet}(a) as a black curve. We detrend the intensity profile by subtracting its smoothed intensity profile, denoted by a red curve in Figure\,\ref{fig:wavelet}(a), using a 6-minute boxcar, and display the residual intensity profile in Figure\,\ref{fig:wavelet}(b) as a blue curve. The detrended intensity profile in Figure\,\ref{fig:wavelet}(b) clearly shows the periodic variations of the intensity. We employ wavelet-analysis technique of \citet{torr98} to retrieve the periodicity in the intensity variations. The standard ``wavelet.pro" routine in the SSW package, where the ``Morlet" function is chosen as the mother function, is applied to the detrended intensity profile with time cadence of 0.2\,minutes and time duration of 420\,minutes. The wavelet power spectrum and global wavelet power spectrum are separately displayed in Figures\,\ref{fig:wavelet}(c) and (d). Here, the confidence levels and global confidence levels are calculated employing the routine ``wave$_{-}$signif.pro" in the SSW package. A significant power with periods in the range of 2.5-7 minutes with a peak period at 4 minutes is evidently identified, lasting for almost 7\,hr in Figure\,\ref{fig:wavelet}(c). More than 110 cycles are identified during the 7\,hr, and no obvious drift of the periods is detected.

\section{Summary and discussion} \label{sec:discussion}

Employing AIA multi-wavelength images on 2012 January 19, reconnection between loops L1 and L2, and coronal condensations in the MD of loops L1 are investigated. During the reconnection, disturbances originating from the reconnection region are detected. They propagate upward across the MD of loops L1 with a mean speed of 200 km\,s$^{-1}$ in the range of 135-265 km\,s$^{-1}$. The speed amplitude ranges in 15-45 km\,s$^{-1}$ with a mean value of 30 km\,s$^{-1}$. The mean speed initially remains at a constant value of 230 km\,s$^{-1}$,  then decreases to 150 km\,s$^{-1}$, and increases to 220 km\,s$^{-1}$ thereafter. This temporal evolution of the mean speed anti-correlates with the light curves of the AIA 131 and 304\,\AA~channels, that clearly show the cooling and condensation process of hotter coronal plasma in the MD of loops L1. It indicates the speed becomes smaller during the condensation process. Two peaks of the histogram of speeds with values of 150 and 210 km\,s$^{-1}$ are identified, separately corresponding to the disturbances originating within and without the condensation process. The disturbances are recurrently generated, with a peak period of 4 minutes in the range of 2.5-7 minutes. 
 
The disturbances propagate upward across the loops L1 that outline the magnetic flux in the corona. In the MD of loops L1, the Alfv\'{e}n speed, $v_{A}$, is estimated using  $v_{A}$=$B$$\times$(4$\pi$\,$n_{p}$\,$m_{p}$)$^{-0.5}$. Here $B$ is the magnetic field strength, $n_{p}$ is the proton number density, and $m_{p}$ is the proton mass. The loops L1 root in the quiet Sun region, $B$ in the MD of loops L1 is hence small, with a value of  $\sim$2\,G obtained from the potential field source surface (PFSS) coronal fields \citep{li18a}. Employing $n_{p}$ of (6$\pm$0.5)$\times$10$^{8}$ cm$^{-3}$ in the quiet Sun region \citep{mcin11} and $B$ of 2$\pm$1 G, $v_{A}$ is calculated to be 180$\pm$95 km\,s$^{-1}$. The sound speed, $C_{s}$, in the MD of loops L1 is also estimated using $C_{s}$=152 $T^{0.5}$, where $C_{s}$ is in km s$^{-1}$, and $T$ is the temperature in MK. As the plasma appears sequentially in AIA 171, 131, and 304\,\AA~channels, we employ $T$ of 0.1-0.9 MK to calculate $C_{s}$, and obtain values of 95$\pm$50 km\,s$^{-1}$. The speed of fast-mode magnetoacoustic waves, $v_{f}$, is 205$\pm$105 km\,s$^{-1}$ using $v_{f}$=$(v_{A}^{2}+C_{s}^{2})^{0.5}$. The observed speeds (200$\pm$65 km\,s$^{-1}$) of disturbances are thus consistent with the estimated $v_{f}$. They are smaller than those reported before \citep{liu11,shen12,yuan13,zhang15}, caused by the smaller Alfv\'{e}n speed in the MD of loops L1 with weaker magnetic field than the larger Alfv\'{e}n speed in the active region loops with stronger magnetic field, but much larger than those of slow-mode magnetoacoustic waves \citep{zhang15}. Moreover, the disturbances are generated quasi-periodically. From these we conclude the quasi-periodic disturbances propagate across the field lines of loops with speeds consistent with those of fast-mode magnetoacoustic waves. Hence the disturbances should represent the QFPM waves. 
 
An increase of the wave speed following a decrease are detected. The wave speed becomes smaller during the condensation process of coronal plasma, resulting in the smaller peak at 150 km s$^{-1}$ of the speed distribution. In the MD of loops L1, coronal condensations lead to the increase of $n_{p}$ and the decrease of $T$, and thus the decrease of $v_{A}$ and $C_{s}$. The decrease of the Alfv\'{e}n speed and sound speed in the propagating path of waves hence causes the decrease of the wave speeds. On the other hand, the QFPM waves may disturb the plasma in the propagating path, thereby causing the coronal condensations in the MD of loops L1.

Only waves propagating upward across the MD of loops L1, rather than other directions, are observed. This may be caused by two reasons: (1) the physical property, e.g., density, magnetic field, and temperature, of the  atmosphere surrounding the reconnection region, as QFPM waves are trapped in regions of higher density, i.e., in regions with a lower Alfv\'{e}n speed \citep{vrsn08,jeli12}; and (2) the employed instrument and wavelength bandpasses \citep{li16c}. 

The energy flux of waves, $E_{w}$, is calculated employing $E_{w}$=0.5\,$n_{p}$\,$m_{p}$\,$v_{amp}$$^{2}$\,$v_{w}$. Here $v_{w}$ is the wave speed, and $v_{amp}$ is the speed amplitude. Using $v_{w}$ of 200$\pm$65 km\,s$^{-1}$ and $v_{amp}$ of 30$\pm$15 km\,s$^{-1}$, $E_{w}$ is measured to be (1.5$\pm$1.4)$\times$10$^{5}$ erg cm$^{-2}$ s$^{-1}$, consistent with those previously reported \citep{liu11,shen13}. The magnetic energy flux, $E_{m}$, during the reconnection is also estimated using $E_{m}$=($B^{2}$\,$v_{in}$)/8$\pi$. Here $v_{in}$ is the reconnection inflowing speed. Employing the moving speeds (5.5$\pm$3.5 km\,s$^{-1}$) of loops L1 toward the reconnection region as $v_{in}$, $E_{m}$ is calculated to be (1.7$\pm$1.6)$\times$10$^{5}$ erg cm$^{-2}$ s$^{-1}$, much smaller than those estimated during flares \citep{asai04,isob05}. Based on these calculations, one then can speculate most of the magnetic energy may be converted to the wave energy through reconnection. This speculation is consistent with the observations that less heating and acceleration of plasma are detected during the reconnection. Moreover, different from \citet{godd16} and \citet{kolo18}, neither the QFPM wave radio signature, nor the reconnection radio signature is detected from Nobeyama observations. This indicates that there is no significant (detectable) non-thermal emission during the reconnection. 

The waves are generated with a peak period of 4 minutes in the range of 2.5-7 minutes. These periods are consistent with the main periods of QFPM waves in \citet{liu11} and \citet{kumar17}, but larger than those in \citet{shen12}, \citet{yuan13}, and \citet{nist14}, during flares. The QFPM waves and their associated flares are suggested to originate from a common source, i.e., the reconnection releasing magnetic energy quasi-periodically during flares \citep{ofman11,shen12}. However, no reconnection process is detected in those flares. In this study, no flare is observed during the reconnection. The QFPM waves directly associated with the reconnection rather than the flare are hence identified.

The QFPM waves originating from reconnection have been theoretically investigated. 
A steady inflow of magnetic flux, e.g., the successive downward motion of loops L1, toward the reconnection region could result in the repetitive and even periodic reconnection \citep{mcla09,murr09}.
As the periodicity of waves, i.e., 4-minute peak period, is the same as that of the chromospheric oscillations, the reconnection may also be modulated by the periodic MHD oscillations \citep{chen06,liu11}. Moreover, plasmoids and reconnection outflows produced in the current sheets during reconnection are theoretically suggested as the exciters of QFPM waves \citep{yang15a,taka16,jeli17}. However, no plasmoid is observed here during the reconnection. More observations and theoretical studies are still needed to understand the physical cause of QFPM waves and their general role in coronal condensations in non-flaring regions.

\acknowledgments

We thank the anonymous referee for many valuable comments that helped us to improve the paper substantially. The authors are indebted to the SDO team for providing the data, and are grateful to Dr. Liping Yang for valuable discussions. The work is supported by the National Foundations of China (11673034, 11533008, 11790304, and 11773039), and Key Programs of the Chinese Academy of Sciences (QYZDJ-SSW-SLH050). L. P. C. received funding from the European Union's Horizon 2020 Research and Innovation Programme under the Marie Sk\l{}odowska-Curie grant agreement No. 707837. This project supported by the Specialized Research Fund for Shandong Provincial Key Laboratory.




\begin{thebibliography}{}

\bibitem[Asai et al. (2004)]{asai04} Asai, A., Yokoyama, T., Shimojo, M., et al.\ 2004, \apj, 611, 557
\bibitem[Aschwanden(2002)]{asch02} Aschwanden, M.\ 2002, \ssr, 101, 1
\bibitem[Chen \& Priest (2006)]{chen06} Chen, P., \& Priest, E.\ 2006, \solphys, 238, 313
\bibitem[Cheng et al. (2018)]{cheng18} Cheng, X., Li, Y., Wan, L., et al.\ 2018, \apj, 866, 64
\bibitem[Goddard et al.(2016)]{godd16} Goddard, C., Nistic\`{o}, G., Nakariakov, V., Zimovets, I., \& White, S.\ 2016, \aap, 594, A96 
\bibitem[Huang et al.(2018)]{huang18} Huang, Z., Xia, L., Nelson, C., et al.\ 2018, \apj, 853, L26
\bibitem[Isobe et al. (2005)]{isob05} Isobe, H., Takasaki, H., \& Shibata, K.\ 2005, \apj, 632, 1184
\bibitem[Jel\'{i}nek et al. (2012)]{jeli12} Jel\'{i}nek, P., Karlick\'{y}, M., \& Murawski, K.\ 2012, \aap, 546, A49
\bibitem[Jel\'{i}nek et al. (2017)]{jeli17} Jel\'{i}nek, P., Karlick\'{y}, M., Van Doorsselaere, T., \& B\'{a}rta, M.\ 2017, \apj, 847, 98
\bibitem[Kolotkov et al. (2018)]{kolo18} Kolotkov, D., Nakariakov, V., \& Kontar, E.\ 2018, \apj, 861, 33
\bibitem[Kumar et al. (2017)]{kumar17} Kumar, P., Nakariakov, V., \& Cho, K.\ 2017, \apj, 844, 149
\bibitem[Lemen et al. (2012)]{lemen12} Lemen, J., Title, A., Akin, D., et al.\ 2012, \solphys, 275, 17
\bibitem[Li et al. (2016a)]{li16a} Li, D., Ning, Z., \& Su, Y.\ 2016a, \apss, 361, 301
\bibitem[Li \& Zhang(2009)]{li09} Li, L., \& Zhang, J.\ 2009, \apj, 703, 877
\bibitem[Li et al.(2016b)]{li16b} Li, L., Zhang, J., Peter, H., et al.\ 2016b, NatPh, 12, 847
\bibitem[Li et al.(2016c)]{li16c} Li, L., Zhang, J., Su, J., \& Liu, Y.\ 2016c, \apj, 829, L33
\bibitem[Li et al. (2018)]{li18a} Li, L., Zhang, J., Peter, H., et al.\ 2018, \apj, 864, L4
\bibitem[Lin et al.(2005)]{lin05} Lin, J., Ko, Y., Sui, L., et al.\ 2005, \apj, 622, 1251
\bibitem[Liu et al.(2010)]{liu10} Liu, R., Lee, J., Wang, T., et al.\ 2010, \apj, 723, L28
\bibitem[Liu et al. (2011)]{liu11} Liu, W., Title, A., Zhao, J., et al.\ 2011, \apj, 736, L13
\bibitem[Liu et al. (2012)]{liu12} Liu, W., Ofman, L., Nitta, N., et al.\ 2012, \apj, 753, 52
\bibitem[Masuda et al.(1994)]{masu94} Masuda, S., Kosugi, T., Hara, H., Tsuneta, S.,  \& Ogawara, Y.\ 1994, Nat, 371, 495
\bibitem[McIntosh et al. (2011)]{mcin11} McIntosh, S., De Pontieu, B., Carlsson, M., et al.\ 2011, \nat, 475, 477
\bibitem[McKenzie(2000)]{mcke00} McKenzie, D.\ 2000, \solphys, 195, 381
\bibitem[McLaughlin et al. (2009)]{mcla09} McLaughlin, J., De Moortel, I., Hood, A., \& Brady, C.\ 2009, \aap, 493, 227
\bibitem[Murray et al. (2009)]{murr09} Murray, M., van Driel-Gesztelyi, L., \& Baker, D.\ 2009, \aap, 494, 329
\bibitem[Ning (2016)]{ning16} Ning, Z.\ 2016, \apss, 361, 22
\bibitem[Nistic\`{o} et al. (2014)]{nist14} Nistic\`{o}, G., Pascoe, D., \& Nakariakov, V.\ 2014, \aap, 569, A12
\bibitem[Ofman et al. (2011)]{ofman11} Ofman, L., Liu, W., Title, A., \& Aschwanden, M.\ 2011, \apj, 740, L33
\bibitem[Pesnell et al. (2012)]{pesn12} Pesnell, W., Thompson, B., \& Chamberlin, P.\ 2012, \solphys, 275, 3
\bibitem[Priest \& Forbes(2000)]{prie00} Priest, E., \& Forbes, T.\ 2000, Magnetic reconnection (Cambridge: Cambridge Univ. Press), 612
\bibitem[Reeves et al.(2015)]{reev15} Reeves, K., McCauley, P., \& Tian, H.\ 2015, \apj, 807, 7
\bibitem[Savage \& McKenzie(2011)]{sava11} Savage, S., \& McKenzie, D.\ 2011, \apj, 730, 98
\bibitem[Shen \& Liu (2012)]{shen12} Shen, Y., \& Liu, Y.\ 2012, \apj, 753, 53
\bibitem[Shen et al. (2013)]{shen13} Shen, Y., Liu, Y., Su, J., et al.\ 2013, \solphys, 288, 585
\bibitem[Shibata (1999)]{shib99} Shibata, K.\ 1999, Astrophys. Spa. Sci., 264, 129
\bibitem[Su et al.(2013)]{su13} Su, Y., Veronig, A., Holman, G., et al.\ 2013, NatPh, 9, 489
\bibitem[Takasao et al.(2012)]{taka12} Takasao, S., Asai, A., Isobe, H., \& Shibata, K.\ 2012, \apj, 745, L6
\bibitem[Takasao \& Shibata (2016)]{taka16} Takasao, S., \& Shibata, K.\ 2016, \apj, 823, 150
\bibitem[Tian et al.(2014)]{tian14} Tian, H., Li, G., Reeves, K., et al.\ 2014, \apj, 797, L14
\bibitem[Torrence \& Compo (1998)]{torr98} Torrence, C., \& Compo, G.\ 1998, Bull. Meteorol. Soc., 79, 61
\bibitem[Tsuneta et al.(1992)]{tsun92} Tsuneta, S., Hara, H., Shimizu, T., et al.\ 1992, \pasj, 44, L63
\bibitem[Vr$\breve{s}$nak \& Cliver (2008)]{vrsn08} Vr$\breve{s}$nak, B., \& Cliver, E.\ 2008, \solphys, 253, 215
\bibitem[Xue et al. (2018)]{xue18} Xue, Z., Yan, X., Yang, L., et al.\ 2018, \apj, 858, L4
\bibitem[Yamada et al.(2010)]{yama10} Yamada, M., Kulsrud, R., \& Ji, H.\ 2010, \rmp, 82, 603
\bibitem[Yan et al.(2018)]{yan18} Yan, X., Yang, L., Xue, Z., et al.\ 2018, \apj, 853, 18
\bibitem[Yang et al. (2015)]{yang15a} Yang, L., Zhang, L., He, J., et al.\ 2015, \apj, 800, 111
\bibitem[Yokoyama et al.(2001)]{yoko01} Yokoyama, T., Akita, K., Morimoto, T., Inoue, K., \& Newmark, J.\ 2001, \apj, 546, L69
\bibitem[Yuan et al. (2013)]{yuan13} Yuan, D., Shen, Y., Liu, Y., et al.\ 2013, \aap, 554, A144
\bibitem[Zhang et al. (2015)]{zhang15} Zhang, Y., Zhang, J., Wang, J., \& Nakariakov, V.\ 2015, \aap, 581, A78

\end{thebibliography}
\end{document}